\documentstyle[amssymb,12pt]{article}

\begin{document}

\title{Killing Vector Fields  in Three Dimensions: A Method to Solve Massive Gravity Field Equations }
\author{ Metin G{\" u}rses\\
{\small Department of Mathematics, Faculty of Sciences}\\
{\small Bilkent University, 06800 Ankara - Turkey}}

\begin{titlepage}
\maketitle
\begin{abstract}
Killing vector fields in three dimensions play important role in
the construction of the related spacetime geometry. In this work
we show that when a three dimensional geometry admits a  Killing
vector field then the Ricci tensor of the geometry is determined
in terms of the Killing vector field and its scalars. In this way
we can generate all products and covariant derivatives at any
order of the ricci tensor. Using this property we give ways of
solving the field equations of Topologically Massive Gravity (TMG)
and New Massive Gravity (NMG) introduced recently. In particular
when the scalars of the Killing vector field (timelike, spacelike
and null cases) are constants then  all three dimensional
symmetric tensors of the geometry, the ricci and einstein tensors,
their covariant derivatives at all orders, their products of all
orders are completely determined  by the Killing vector field and
the metric. Hence the corresponding three dimensional metrics are
strong candidates of solving all higher derivative gravitational
field equations in three dimensions.

\end{abstract}
\end{titlepage}

 \maketitle

\section{Introduction}

Einstein gravity in three dimensions is trivial because it
contains no dynamics. A theory in three dimensions with dynamics
is known as topologically massive gravity (TMG)
\cite{djt},\cite{djt1}. The action of TMG is the sum of the
standard Einstein-Hilbert and Chern-Simons terms in three
dimensions. TMG with cosmological constant is considered  recently
\cite{car}, \cite{song}. This theory violates parity, it is
renormalizable but not unitary. Recently a new massive gravity
theory (NMG) is introduced. In the framework of perturbation
theory NMG is both renormalizable and unitary
\cite{bht}-\cite{bht2} (see also \cite{des0}, \cite{bay}).
Although higher derivative theories play important role to have a
consistent quantum gravity  it is however very difficult to obtain
classical solutions of these theories. For this purpose there have
been several attempts to solve these theories, in particular to
find black hole solutions. In these attempts different methods
were used. In \cite{gur4} we used perfect fluid solutions of
Einstein theory to solve the TMG field equations. In \cite{clem0}
- \cite{clem5} by assuming commuting two Killing vector fields
Clement gave some solutions of the TMG and NMG field equations.
Another method is the principle of symmetric criticality
\cite{pal}, \cite{tor}, \cite{des1}. In \cite{sez1} algebraic
classification of the ricci tensor is used and reviewed all known
solutions of the TMG. In this work it has been also emphasized the
importance of the Killing vector fields in obtaining the solutions
of the TMG field equations. It was shown that (see also
\cite{annin}) all solutions are locally belong to one the three
solutions: Timelike-squashed $AdS_{3}$, Spacelike-squashed
$AdS_{3}$ and AdS pp-waves  \cite{des4}, \cite{Ay1}, \cite{Ay2}.
In \cite{sez3} Chow et al used the Kundt metrics to solve TMG
field equations and they obtained new solutions of the TMG.
Another method is to use the G{\" o}del type metrics
\cite{gur1}-\cite{gur3} in three dimensions \cite{gur5}. There are
also some  other attempts \cite{bar2}-\cite{vuor} to solve the
field equations of TMG. AdS pp-wave solutions of NMG filed
equations were considered in \cite{Ay3}. The importance of the
Killing vectors in finding the solutions of the massive gravity
field equations in three dimensions is also observed in some of
these works. If they exist, Killing vectors in three dimensions
are not only important in the derivation of the exact solutions
but they control the whole geometry of spacetime.

In this work we shall introduce a new method for three dimensional
gravity theories. The starting point is to assume a one parameter
family of isometry. This implies  existence of a Killing vector
field in the spacetime geometry. In general this assumption may
simplify and reduce the number the field equations by choosing a
suitable coordinate patch so that one coordinate in the line
element becomes cyclic. In the case of four and higher dimensional
geometries this may give information about some components of the
curvature and hence the ricci tensors but it does not provide all
components of these tensors. We shall show that this is possible
in three dimensions.

If a three dimensional spacetime admits a non-null Killing vector
field, writing it in terms of two scalars, its norm and rotation
scalar, we show that the metric and the ricci tensors are
completely determined in terms of the Killing vector, its scalars
and vectors obtained through these scalars (Theorem 1). Physical
interpretation of the corresponding spacetime metric is that it
solves the Einstein-Maxwell- Dilaton field equations in three
dimensions. When the scalars of the Killing vector field  become
constant then the corresponding ricci tensor becomes relatively
simple (Theorem 2). We show that the spacetime metric, in this
case, solve the field equations of TMG and NMG. Indeed these
metrics (for timelike and spacelike cases) solve not only these
field equations, but they solve all higher (curvature) derivative
theories at any order. When the Killing vector field is null we
have similar result (Theorem 3). Without referring to the metric
tensor we construct the Einstein tensor when the spacetime
geometry admits a null vector. The corresponding metric solves
Einstein field equations with a null fluid distribution and an
exotic scalar field. When rotation  scalar of the null Killing
vector field become constant the Einstein tensor becomes more
simpler (Theorem 4) and the metric solves the field equations of
TMG and NMG. Both for non-null and null cases when the scalars of
the Killing vector fields become constant then the metric solves
not only the field equations of TMG and NMG  but  solves also all
higher derivative gravitational filed equations in three
dimensions .

In Section 2 we give the ricci and metric tensors in terms of the
non-null Killing vector field and its scalars. The Einstein tensor
represents a charged fluid with a dilaton field.  In particular if
the scalars are constants the ricci tensor is expressed in terms
of the Killing vector field and he Gaussian curvature of the two
dimensional background space in Section 3. In Section 4 we show
that when the Killing vector field is null then the ricci tensor
is calculated in terms of the null Killing vector filed and its
rotation scalar. In Section 5 we investigate the  the case when
the rotation scalar is a constant. In Sections 6 and 7, by using
the ricci tensors found in the previous Sections, we solve the
field equations of the TMG and NMG respectively.

\section{Non-Null Killing Vectors in Three Dimensions}

In this section we shall construct the metric and Einstein tensors
in terms of scalar functions of a non-null Killing vector. Our
conventions in this paper are similar to the convention of
Hawking-Ellis  \cite{he}.

Let $u^{\mu}$ be a non-null Killing vector field satisfying the
Killing equation

\begin{equation}\label{eq11}
u_{\mu \, ; \, \nu}+u_{\nu \, ; \, \mu}=0
\end{equation}
with $u_{\alpha}\, u^{\alpha} =\alpha \ne 0$. One can show that

\begin{equation}\label{eq12}
u^{\alpha}\, u_{\alpha \, ; \, \mu}={1 \over 2}\,
\alpha_{,\mu},~~u^{\alpha}\, u_{\mu \, ; \, \alpha }=-{1 \over
2}\, \alpha_{,\mu}
\end{equation}
The norm function $\alpha$ is in general not a constant. It is
possible to write the Killing equation as

\begin{equation}\label{eq13}
u_{\mu \, ; \, \nu}={1 \over 2}\, (u_{\mu \, ; \, \nu}-u_{\nu \, ;
\, \mu})=\eta_{\mu \nu \alpha}\, v^{\alpha}
\end{equation}
where $v^{\alpha}$ is  any vector field and $\eta_{\mu \nu
\alpha}$ is the Levi-Civita alternating symbol. Using (\ref{eq12})
we find that

\begin{equation}\label{eq14}
v^{\rho}=w u^{\rho}+\eta^{\rho \mu \sigma}\, \phi_{,\mu}\,
u_{\sigma}
\end{equation}
where $w={u_{\mu}v^{\mu} \over \alpha}$. Using (\ref{eq14}) in
(\ref{eq13}) we get

\begin{equation}\label{eq16}
u_{\mu \, ; \, \nu}=w\,\eta_{\mu \nu \alpha}\,
u^{\alpha}+u_{\mu}\, \phi_{,\nu}-u_{\nu}\, \phi_{,\mu}
\end{equation}
where $\phi={1 \over 2}\, \ln \alpha$. The functions  $w$ and
$\phi$ specify the Killing vector field and they are assumed to be
linearly independent. Differentiating (\ref{eq16}) with respect to
$x^{\sigma}$ and anti-symmetrizing both sides with respect $\nu$
and $\sigma$ we get

\begin{eqnarray}\label{eq17}
u_{\mu \, ; \, \nu \, \sigma}-u_{\mu \, ; \, \sigma \, \nu}&=&
w_{,\sigma}\,\eta_{\mu \nu \alpha}\,u^{\alpha}-w_{,\nu}\,
\eta_{\mu \sigma \alpha}\,u^{\alpha}-w\, \eta_{\mu \nu
\alpha}\,\phi^{\alpha}\, u_{,\sigma}+w\,\eta_{\mu \sigma
\alpha}\,\phi^{\alpha}\, u_{\nu} \nonumber \\
&&- 2\,\eta_{\nu \sigma \alpha}\,u^{\alpha}\,
\phi_{,\mu}-u_{\nu}\, \phi_{,\sigma}\, \phi_{, \mu}+u_{\sigma}\,
\phi_{,\mu}\, \phi_{,\nu}-u_{\nu}\, \phi_{;\mu
\sigma}+u_{\sigma}\, \phi_{;\mu \nu} \nonumber \\
&&+w^2\,(g_{\mu \nu}\, u_{\sigma}-g_{\mu \sigma}\, u_{\nu})
\end{eqnarray}
Using the identities

\begin{eqnarray}\label{eq18}
w_{,\sigma}\,\eta_{\mu \nu \alpha}\,u^{\alpha}+w_{,\mu}\,\eta_{
\nu \sigma \alpha}\,u^{\alpha}+w_{,\nu}\,\eta_{\sigma \mu
\alpha}\,u^{\alpha}=\eta_{\sigma \mu \nu}\,u^{\alpha}\, \phi_{,\alpha}=0,\\
\eta_{\mu \nu \alpha}\,\phi^{\alpha}\,u_{,\sigma}+\eta_{ \nu
\sigma \alpha}\,\phi^{\alpha}\,u_{,\mu}+\eta_{\sigma \mu
\alpha}\,\phi^{\alpha}\,u_{,\nu}= \eta_{\sigma \mu \nu}\,
\phi^{\alpha}\, u_{\alpha}=0
\end{eqnarray}
then Eq. (\ref{eq17}) reduces to (Here $\phi^{\alpha}=g^{\alpha
\mu}\,\phi_{,\mu}$)

\begin{eqnarray}\label{eq19}
u_{\mu \, ; \, \nu \, \sigma}-u_{\mu \, ; \, \sigma \, \nu}&= &w\,
\eta_{\nu \sigma \alpha}\, \phi^{\alpha}\, u_{\mu}-\eta_{\nu
\sigma \alpha}\, u^{\alpha}\, y_{\mu}+\phi_{,\mu}\, (u_{\sigma}\,
\phi_{,\nu}-u_{\nu}\, \phi_{,\sigma}) \nonumber \\
&&+w^2\, (g_{\mu \nu}\, u_{\sigma}-g_{\mu \sigma}\, u_{\nu})
\end{eqnarray}
where

\begin{equation}
y_{\mu}=w_{,\mu}+2w\, \phi_{,\mu}
\end{equation}
Using the ricci identity $u_{\mu \, ; \, \nu \, \sigma}-u_{\mu \,
; \, \sigma \, \nu}=R^{\rho}\,_{\mu \nu \sigma}\, u_{\rho}$ in
(\ref{eq19}) we get

\begin{eqnarray}\label{eq20}
R^{\rho}\,_{\mu \nu \sigma}\, u_{\rho}&= &w\, \eta_{\nu \sigma
\alpha}\, \phi^{\alpha}\, u_{\mu}-\eta_{\nu \sigma \alpha}\,
u^{\alpha}\, y_{\mu}+\phi_{,\mu}\, (u_{\sigma}\,
\phi_{,\nu}-u_{\nu}\, \phi_{,\sigma}) \nonumber \\
&&+w^2\, (g_{\mu \nu}\, u_{\sigma}-g_{\mu \sigma}\, u_{\nu})
\end{eqnarray}
and hence we obtain

\begin{equation}
R^{\rho}\,_{\nu}\, u_{\rho}=-\zeta_{\nu}-(\nabla^2\, \phi+2 w^2)\,
u_{\nu}
\end{equation}
where
\begin{equation}
\zeta_{\mu}=\eta_{\mu \alpha \beta}\, y^{\alpha}\, u^{\beta}
\end{equation}
In three dimensions the Weyl tensor vanishes and hence curvature
tensor is expressed solely in terms of the ricci tensor

\begin{equation}\label{eq21}
R_{\rho \mu \nu \alpha}=-R_{\mu \nu}\, g_{\rho \alpha}+R_{\rho
\nu}\, g_{\mu \alpha}+R_{\mu \alpha}\, g_{\rho \nu}-R_{\rho
\alpha}\, g_{\mu \nu}+{R \over 2}\,(g_{\rho \alpha}\, g_{\mu
\nu}-g_{\rho \nu}\, g_{\mu \alpha})
\end{equation}
Combining equations (\ref{eq20}) and (\ref{eq21}) we obtain

\begin{equation}\label{eqn22}
\bar{R}_{\mu \nu}\, u_{\sigma}-\bar{R}_{\mu \sigma}\,
u_{\nu}+\zeta_{\sigma}\, g_{\mu \nu}-\zeta_{\nu}\, g_{\mu
\sigma}=-\eta_{\nu \sigma \alpha}\,u^{\alpha}\, y_{\mu}+w\,
\eta_{\nu \sigma \alpha}\, \phi^{\alpha}\, u_{\mu},\\
\end{equation}
where
\begin{eqnarray}
 &\bar{R}_{\mu \nu}=-R_{\mu \nu}+(P+{R \over 2}-w^2)\, g_{\mu
\nu}-\phi_{;\mu \nu}-\phi_{,\mu}\, \phi_{,\nu},\\
&P=\nabla^2\, \phi+2 \phi_{, \alpha}\, \phi^{\alpha}+2 w^2
\end{eqnarray}

In addition to the identities (\ref{eq18}) we have also the
following identities

\begin{eqnarray}
z_{\mu}\, u_{\sigma}-z_{\sigma}\, u_{\nu}&=& \alpha \,\eta_{\mu
\sigma \rho}\, \phi^{\rho}, \label{ozd1}\\
\zeta_{\mu}\, y_{\sigma}-\zeta_{\sigma}\, y_{\nu}&=&
-(y_{\alpha}\,y^{\alpha}) \,\eta_{\mu \sigma \rho}\, u^{\rho}
\label{ozd2}
\end{eqnarray}
Using these identities in (\ref{eq22}) we arrive at the following
theorem:

\vspace{0.5cm}

\noindent {\bf Theorem 1}. {\it If a three dimensional spacetime
admits  a non-null Killing vector field $u^{\mu}$ then its metric
and the Einstein tensors are respectively given by

\begin{eqnarray}
g_{\mu \nu}& =&{1 \over \alpha}\, u_{\mu}\, u_{\nu}+{1 \over
\zeta^{\alpha}\,\zeta_{\alpha}}\, \zeta_{\mu}\, \zeta_{\nu}+{1
\over y_{\alpha}y^{\alpha}}\, y_{\mu}\, y_{\nu}, \label{eq22}\\
G_{\mu \nu}&=&-{H \over \alpha}\, u_{\mu}\, u_{\nu}+{1 \over
\alpha}\,[(w\, z_{\mu}-\zeta_{\mu})\, u_{\nu}+(w\,
z_{\nu}-\zeta_{\mu})\, u_{\mu}] \label{eq23} \\
&&+(\nabla^2 \, \phi+w^2)\, g_{\mu \nu}-\phi_{,\mu}\,
\phi_{,\nu}-\phi_{; \mu \nu}, \\
H&=&2 \nabla^2 \phi+3w^2+{1 \over 2}\,R -\phi^{\alpha}\,
\phi_{,\alpha},\\
z_{\mu}&=&\eta_{\mu \alpha \beta}\, \phi^{\alpha}\, u^{\beta}
\end{eqnarray}
where $R$ is the ricci scalar provided that the functions $w$ and
$\phi$ are not constant at the same time. Here we assume that
$\alpha$ nowhere vanishes in the spacetime.}

\vspace{0.5cm}

\noindent {\bf Remark 1}. When $u^{\mu}$ is a timelike Killing
vector field, letting $\alpha=-e^{2 \phi}$, then it can be shown
that the source of the Einstein equations is composed of a charged
fluid distribution and a dilaton field $\phi$ (in Einstein frame).
The electromagnetic field tensor $f_{\mu \nu}$ is given by

\begin{equation}\label{eq24}
f_{\mu \nu}=w\,\eta_{\mu \nu \alpha}\, u^{\alpha}-u_{\mu}\,
\phi_{,\nu}+u_{\nu}\, \phi_{,\mu}
\end{equation}
so that

\begin{equation}
f^{\mu \nu}\, _{;\mu}=-\zeta^{\nu}+(\nabla^2 \, \phi-2 w^2)\,
u^{\nu}
\end{equation}
The Maxwell energy momentum tensor is given as

\begin{eqnarray}\label{eq25}
M_{\mu \nu}&=&-{1 \over 2} \alpha\, (w^2+\phi_{,\alpha}\,
\phi^{\alpha})\, g_{\mu \nu}+(w^2+\phi_{,\alpha}\,
\phi^{\alpha})\,u_{\mu}\,u_{\nu} \nonumber \\
&&+\alpha\, \phi_{,\mu}\, \phi_{, \nu}-w \, (z_{\mu}\,
u_{\nu}+z_{\nu}\, u_{\mu})
\end{eqnarray}
Then field equations are

\begin{eqnarray}
G_{\mu \nu}&=&{1 \over 4}\, e^{-2 \phi}\, M_{\mu \nu}+e^{-2
\phi}\,(-2 \nabla^2 \phi+4\,\phi_{, \alpha}\, \phi^{\alpha}
+e^{-2\phi}\,f^2-{R \over 2}) \, u_{\mu}\, u_{\nu}
\nonumber\\
&&+(\nabla^2 \phi+{1 \over 4}\, e^{-2 \phi}\, f^2) g_{\mu \nu}
-e^{-2 \phi}\, (J_{\mu}\, u_{\nu}+J_{\nu}\, u_{\mu})-\phi_{; \mu
\nu}
\end{eqnarray}
where $J^{\,\mu}=f^{\alpha \mu}\, _{;\alpha}$ and $f^2 =f_{\alpha
\beta}\, f^{\alpha \beta}$.

\vspace{0.3cm}

All our analysis so far is valid when the norm $\alpha$ of the
Killing vector field $u^{\mu}$ is different than zero everywhere
in spacetime . If there are points where $\alpha=0$ we exclude
them from the spacetime manifold for causality reasons. This set
of points define the Killing horizons which are in general
coordinate singularities of the spacetime \cite{he}. We may find
suitable coordinate patches where these coordinate singularities
are removed (maximal extension) \cite{cart}. Examples to such
Killing horizons in three dimensions are discussed in the BTZ
spacetime \cite{teit}. Killing horizon, in general,  is a subset
of the set of fixed points of the isometry whose generator is the
Killing vector $u$. Such a set is called local isometry horizon
(LIH) by Carter \cite{cart}. Fixed points of an isometry are the
set of points where the Killing vector field vanish, i.e.,
critical points of $u$, $u(p)=0$. It is obvious that our analysis
fails at the points of LIH. For a given Killing vector field we
have \cite{eisen}

\begin{equation}\label{eis}
u_{\mu;\alpha \beta}=u_{\rho}\, R^{\rho}\,_{\beta \mu \alpha}
\end{equation}
At the critical points of $u$ the second covariant derivative of
the Killing vector field vanishes. This implies that all higher
derivatives of $u$ at $p$ are either zero or expressed in terms of
the first partial derivative $F_{\mu \nu}(p)=u_{\mu,\nu}(p)$. The
integrability of (\ref{eis}) at $p$ gives

\begin{equation}
F_{\mu \alpha}(p)\, R^{\alpha}\,_{\nu}+F_{\nu \alpha}(p)\,
R^{\alpha}\,_{\mu}=0
\end{equation}
which imply that the Segre type of the ricci tensor is special at
the critical points of $u$ \cite{hall}, \cite{carot}. In a
neighborhood $U$ of the critical point $p$ the Killing vector has
the linearized form

\begin{equation}
u_{\mu}=F_{\mu \alpha}(p)\,Z^{\alpha},
\end{equation}
where $Z^{\mu}=x^{\mu}-p^{\mu}$. Theorem 1 is valid in $U$ with

\begin{equation}
\alpha=F^{\mu}\,_{\alpha}(p)\, F_{\mu \beta}(p)\,
Z^{\alpha}\,Z^{\beta},~ w=-{1 \over 2\, \alpha}\, \eta^{\mu \nu
\alpha}\, F_{\mu \nu}(p)\, F_{\alpha \beta}(p)\, Z^{\beta}.
\end{equation}
To identify the real structure of the spacetime about the critical
points for a given particular exact solution it may be useful to
study the maximal extension of the corresponding coordinate
patches.

If the norm $\alpha$ of a Killing vector  field is constant
everywhere in spacetime then there are no fixed points of the
corresponding isometry. Then Theorem 1 is valid everywhere in
spacetime. This is the case which will be considered in the next
sections. This will be true not only for spacelike and timelike
Killing vector fields but true also for the null  Killing vector
fields.

\section{Non-null Case With Constant Scalars}

In the previous section we assumed that the scalars  $w$ and
$\phi$ of the Killing vector field are not simultaneously
constants. If $w$ and $\phi$ are both constants then we get

\begin{equation}\label{eq26}
u_{\mu \, ; \, \nu}=w\,\eta_{\mu \nu \alpha}\, u^{\alpha}
\end{equation}
Then following the above procedure and using (\ref{eq23}) we have
the following result.

 \vspace{0.5cm}

\noindent {\bf Theorem 2}. {\it If a three dimensional spacetime
admits a non-null Killing vector field $u^{\mu}$ with $w$ and
$\alpha$ are both constants then the Einstein tensor is given as

\begin{equation}\label{eqn27}
G_{\mu \nu}=-{1 \over \alpha}\,(3 w^2+{1 \over 2}\,R)\, u_{\mu}\,
u_{\nu}+w^2\, g_{\mu \nu}
\end{equation}
}

\vspace{0.5cm}

\noindent In this case, since the vector $\zeta_{\mu}$ vanishes in
(\ref{eqn22}), the metric tensor is not obtained.

\vspace{0.4cm}

\noindent {\bf Remark 2}.
 If $u^{\mu}$ is a timelike Killing vector field the
above Einstein tensor represents a dust distribution in a
spacetime with a positive cosmological constant.

\vspace{0.5cm}

\noindent
 For completeness let us give the spacetime metrics of
each case: We shall consider the case where $\alpha$ and $w$ are
constants.

\vspace{0.2cm}

\noindent {\bf 1}. Let $u^{\mu}=-{1 \over
u_{0}}\,\delta^{\mu}_{0}$ be the timelike vector field then
$u^{\mu}\, u_{\mu}=-1$. Hence $\alpha=-1$ and $u_{\mu}=g_{\mu
\alpha}\, u^{\alpha}$, then $g_{00}=-u_{0}^2, g_{01}=-u_{0}\,
u_{1}, g_{02}=-u_{0}\,u_{2}$. Then the spacetime metric can be
taken as ($x^{\mu}=(t,r,z)$)

\begin{equation}\label{god1}
ds^2=-(u_{0}\, dt+u_{1}\, dr + u_{2}\, dz)^2+M^2 \,dr^2+2L dr
dz+N^2\, dz^2
\end{equation}
where $M^2=g_{11}+u_{1}^2, N^2=g_{22}+u_{2}^2, L=g_{12}+u_{1}\,
u_{2}$. Here $u_{0}$ is a constant (due to the Killing equation)
and the metric functions $M$,$N$, $L$ , $u_{1}$ and $u_{2}$
depend on the variables $r$ and $z$. The metric in (\ref{god1}) is
of G{\" o}del type which was used in \cite{gur5}. The Einstein
tensor is given in (\ref{eqn27}) with $\alpha=-1$. The only field
equation is

\begin{equation}
u_{2,r}=u_{1,z}+2w \,\Delta/u_{0},\label{kil00}
\end{equation}
where $\Delta=|u_{0}| \sqrt{M^2\, N^2-L^2}$. For simplicity
consider the case $L=0$. Then $R=2K+2w^2$ where $K$ is the
Gaussian curvature of the locally Euclidian two dimensional
background space with the line element $ds_{2}^2=M^2 \,dr^2+N^2\,
dz^2$. Metric (\ref{god1}) gives the Einstein tensor (\ref{eqn27})
where the only field equation is given in equation (\ref{kil00})
with $\alpha=-1$ and $w$ is constant.

\vspace{0.3cm}

\noindent {\bf 2}. Let  $u^{\mu}={1 \over
u_{2}}\,\delta^{\mu}_{2}$ be the spacelike vector field then
$u^{\mu}\, u_{\mu}=1$. Hence $\alpha=1$ and $u_{\mu}=g_{\mu
\alpha}\, u^{\alpha}={1 \over u_{2}}\, g_{\mu 2}$, then
$g_{22}=u_{2}^2, g_{12}=u_{1}\, u_{2}, g_{02}=u_{0}\,u_{2}$. Then
spacetime metric can be taken as

\begin{equation}\label{god2}
ds^2=(u_{2}\, dz+u_{1}\, dr+u_{0}\, dt)^2+M^2 \,dr^2+2L \,dr
dt-N^2 \,dt^2,
\end{equation}
where $M^2=g_{11}-u_{1}^2$, $N^2=u_{0}^2-g_{00},
L=g_{01}-u_{0}\,u_{1}$, $u_{2}$ is a constant (due to the Killing
equation) and the metric functions $M$,$N$, $L$, $u_{1}$ and
$u_{0}$ depend on the variables $t$ and $r$. The Einstein tensor
is given in (\ref{eqn27}) with $\alpha=1$. The only field equation
is

\begin{equation}
u_{0,r}=u_{1,t}+2\, w \,\Delta/u_{2},\label{kil01}
\end{equation}
where $\Delta=|u_{2}|\, \sqrt{M^2\, N^2+L^2}$. In both cases the
ricci scalar is $R$ is not  constant. In this case as well, take
$L=0$ for simplicity then we have $R=2K+2w^2$ where $K$ is the
Gaussian curvature of two dimensional locally minkowskian
background space with the line element $ds_{2}^2=M^2 \,dr^2-N^2
\,dt^2$. Metric (\ref{god2}) gives the Einstein tensor
(\ref{eqn27}) where the only field equation is given in
(\ref{kil01}) with $\alpha=1$ and $w$ constant. In both cases the
ricci and Einstein tensors are respectively given as
\begin{eqnarray}
R_{\mu \nu}&=&-{1 \over \alpha}\,(4 w^2+K)\, u_{\mu}\,
u_{\nu}+(2w^2+K)\, g_{\mu \nu}, \label{eqn270} \\
G_{\mu \nu}&=&-{1 \over \alpha}\,(4 w^2+K)\, u_{\mu}\,
u_{\nu}+w^2\, g_{\mu \nu}. \label{eqn271}
\end{eqnarray}

\noindent It is possible to pass from timelike case to spacelike
case by complex transformation \cite{clem2}. When the background
two dimensional spaces are de-sitter or anti de-sitter, in each
case, these solutions are known as timelike squashed $AdS_{3}$ and
spacelike squashed $AdS_{3}$ respectively
\cite{sez1},\cite{annin}, \cite{clem0}.

\vspace{0.5cm}

When $R$ is a constant, the Cotton tensor and the Laplacian of the
ricci tensor take simple forms (for both cases). They will used
later.

\begin{eqnarray}
C_{\mu \nu}&=&-w\,(4 w^2+K)[g_{\mu \nu}-{3 \over
\alpha}\, u_{\mu}\, u_{\nu}], \\
\nabla^2 R_{\mu \nu}&=&-2 w^2\,(4 w^2+K)[g_{\mu \nu}-{3 \over
\alpha}\, u_{\mu}\, u_{\nu}]
\end{eqnarray}

\section{Three Dimensional Spacetime Admitting a Null Killing
Vector Field}

In this section we shall construct the Einstein tensor with
respect to the scalar functions of  a null Killing vector field.

Let $\xi_{\mu}$ be Killing Vector field. It satisfies the Killing
equation

\begin{equation}\label{kil1}
\xi_{\mu;\,\nu}={1 \over 2}
(\xi_{\mu;\,\nu}-\xi_{\nu;\,\mu})=\eta_{\mu \nu \alpha}\,
v^{\alpha}
\end{equation}
where $\eta_{\mu \nu \alpha}=\sqrt{-det(g)}\, \epsilon_{\mu \nu
\alpha}$ and $v^{\mu}$ is an arbitrary vector field and a
semicolon defines the covariant derivative with respect to the
metric $g_{\mu \nu}$. Here the $\epsilon_{\mu \nu \alpha}$ is the
three dimensional Levi-Civita alternating symbol. Since
$\xi^{\alpha}\, \xi_{\alpha;\beta}=0$ and $\xi^{\alpha}\,
\xi_{\beta;\alpha}=0$, then $v^{\mu}=w \xi^{\mu}$. Hence
(\ref{kil1}) becomes

\begin{equation}\label{kil2}
\xi_{\mu;\,\nu}= w \,\eta_{\mu \nu \alpha}\, \xi^{\alpha}
\end{equation}
where $w$ is an arbitrary function. Taking one more covariant
derivative of (\ref{kil2}) we find that

\begin{equation}\label{kil3}
\xi_{\mu;\, \nu \alpha}=w_{,\alpha}\, \eta_{\mu \nu \rho}\,
\xi^{\rho}-w^2\, (g_{\mu \alpha}\, \xi_{\nu}-g_{\nu \alpha}\,
\xi_{\mu})
\end{equation}
Using this equation and the Ricci identity we get

\begin{eqnarray}\label{kil4}
\xi_{\mu;\, \nu \alpha}-\xi_{\mu; \,\alpha \nu}&=&w_{,\alpha}\,
\eta_{\mu \nu \rho}\, \xi^{\rho}-w_{,\nu}\, \eta_{\mu \alpha
\rho}\, \xi^{\rho}-w^2\,(g_{\mu \alpha}\, \xi_{\nu}-g_{\nu \mu}\,
\xi_{\alpha}) \nonumber\\
&=&R^{\rho}\, _{\mu \nu \alpha}\, \xi_{\rho}.
\end{eqnarray}
On the other hand, since the Weyl tensor vanishes in three
dimensions the Riemann tensor is expressed totally in terms of the
ricci tensor as expressed in (\ref{eq21}). Using (\ref{kil4}) and
(\ref{eq21}) we find

\begin{equation}\label{id0}
Q_{\mu \nu}\, \xi_{\alpha}-Q_{\mu \alpha}\,
\xi_{\nu}-\zeta_{\nu}\, g_{\mu \alpha}+\zeta_{\alpha}\, g_{\mu
\nu}=w_{,\alpha}\, \eta_{\mu \nu \rho}\,
\xi^{\rho}-w_{,\nu}\,\eta_{\mu \alpha \rho}\,\xi^{\rho}
\end{equation}
where
\begin{eqnarray}\label{id1}
\zeta_{\mu}&=&\eta_{\mu \alpha \beta}\, w^{\alpha}\, \xi^{\beta},\\
Q_{\mu \nu}&=&-R_{\mu \nu}+(w^2+{R \over 2})\, g_{\mu \nu}.
\end{eqnarray}
Here $R$ is the ricci scalar and $w^{\alpha}=g^{\alpha
\mu}\,w_{,\mu}$. Let a vector $X_{\mu}$ be defined as

\begin{equation}
X_{\mu}=\eta_{\mu \alpha \beta}\, F^{\alpha}\, \xi^{\beta}
\end{equation}
where $F^{\mu}=g^{\mu \alpha}\, F_{,\alpha}$ and $F$ is any
function independent of $t$. It is easy to show that $\xi^{\mu}\,
X_{\mu}=X^{\mu}\,X_{\mu}=0$. Since the orthogonal null vectors can
only be parallel then we have

\begin{equation}
X_{\mu}=\sigma \xi_{\mu},
\end{equation}
Hence

\begin{equation}\label{psi}
\zeta_{\mu}=\psi \xi_{\mu},
\end{equation}
and
\begin{equation}\label{id2}
\xi_{\mu}\, w_{,\nu}-\xi_{\nu}\, w_{,\mu}= \psi\,\eta_{\mu \nu
\alpha}\, \xi^{\alpha}, ~~ \psi \ne 0
\end{equation}
which leads also

\begin{equation}
w^{\alpha}\,w_{,\alpha}=\psi^2.
\end{equation}
Using (\ref{id2}) in  (\ref{id0}) we obtain

\begin{equation}
Q_{\mu \nu}+\psi \, g_{\mu \nu}-{1 \over \psi}\, w_{,u}\,
w_{,\nu}+\rho \xi_{\mu}\, \xi_{\nu}=0.
\end{equation}
Hence

\begin{equation}\label{ricci}
R_{\mu \nu}=\rho \, \xi_{\mu}\, \xi_{\nu}-(\psi+2 w^2)\, g_{\mu
\nu}-{1 \over \psi}\, w_{,u}\, w_{,\nu}
\end{equation}
where $\rho$ is a function to be determined and the ricci scalar
is given by

\begin{equation}
R=-6\, w^2-4\psi.
\end{equation}
Then we have the following theorem:

\vspace{0.5cm}

\noindent {\bf Theorem 3}. {\it Let $\xi^{\mu}$ be a null Killing
vector field of  a three dimensional spacetime, then the
corresponding Einstein tensor is given by

\begin{equation}\label{ein}
G_{\mu \nu}= \rho\, \xi_{\mu}\, \xi_{\nu}-{1 \over \psi}\,
w_{,u}\, w_{,\nu}+(w^2+\psi)\, g_{\mu \nu}
\end{equation}
where $\rho$ is a function to be determined. Furthermore the
scalar field satisfies the following partial differential equation

\begin{equation}\label{scal}
\nabla_{\mu}\,({1 \over \psi}\, w^{\mu})=2w.
\end{equation}}

\vspace{0.5cm}

\noindent It is quite interesting that by imposing a null Killing
vector in a three dimensional spacetime geometry we obtain the
form of the Einstein tensor without knowing the metric tensor
explicitly, except the function $\rho$. The source of the Einstein
field equations is composed of a null fluid and a scalar field.
The scalar field satisfies a nonlinear differential equation
(\ref{scal}). From the Einstein equations the energy momentum
tensor $T_{\mu \nu}$ of the source satisfies $T_{\mu \nu}\,
\xi^{\mu}\, \xi^{\nu}=0$. For any  timelike unit vector field
$u^{\mu}$ we have $T_{\mu \nu}\, u^{\mu}\, u^{\nu}=\rho
\,(\xi^{\mu}\, u_{\mu})^2-{1 \over \psi}\,(w_{,\mu}\,
u^{\mu})^2+(\psi+w^2)$. Choosing $\psi=-\sqrt{w_{,\mu}\, w^{\mu}}$
then the sign of the energy depends on the sign of the function
$\rho$.

As a summary we can conclude in this section as: If $\xi_{\mu}$ is
a null Killing vector field of a three dimensional spacetime
geometry then the corresponding metric solves the Einstein-null
fluid-scalar field equations.The only field equations are
(\ref{kil2}) and the scalar field equation given in
Eq.(\ref{scal}).

\vspace{0.5cm}

\section{Null Case Without Scalar Field}

In the previous section we considered the case $\psi \ne 0$.
Vanishing of $\psi$ is a distinct case. Due to this reason we
consider it in a separate section. We have the following result:

\vspace{0.3cm}

\noindent {\bf Theorem 4}.\,{\it When the scalar function $\psi$
in Theorem 3 is zero then $w$ becomes a constant and the ricci,
Einstein tensors take simple forms

\begin{eqnarray}
R_{\mu \nu}&=&\rho \, \xi_{\mu}\, \xi_{\nu}-2 w^2\, g_{\mu
\nu},\label{ric0}\\
G_{\mu \nu}&=& \rho\, \xi_{\mu}\, \xi_{\nu} +w^2\, g_{\mu \nu},
\label{ein}
\end{eqnarray}
with $R=-6w^2$.}

\vspace{0.3cm}

\noindent
 For explicit solutions we shall adopt
some simplifying assumptions.

\vspace{0.3cm}

\noindent
 {\bf (5.A)}. As an example let us assume that
$\xi^{\mu}=\delta^{\mu}_{0}$. Then $\xi_{\mu}=g_{\mu \alpha}\,
\xi^{\alpha}=g_{\mu 0}$. Since $\xi_{\mu}$ is a null vector then
$\xi_{0}=0$. Hence we get $g_{00}=0, g_{01}=\xi_{1},
g_{02}=\xi_{2}$. The corresponding spacetime metric can be given
as ($x^{\mu}=(t,r,z)$)

\begin{equation}
ds^2=2 (\xi_{\mu}\, dx^{\mu})\, dt+m^2\, dr^2+n^2 dz^2+2\,\ell\,
dr dz
\end{equation}
where $\xi_{1}, \xi_{2}$, $m$, $n$ and $\ell$ are functions of $r$
and $z$. Let $\ell=0$, then $ \det(g) \equiv
-\Delta^2=-(\xi_{1}^2\, n^2+\xi_{2}^2\, m^2)$, and the function
$\psi$ defined in (\ref{psi}) takes the form  ~$\psi={\xi_{1}\,
w_{,z}-\xi_{2}\, w_{,r} \over \Delta}$  and

\begin{equation}
g_{\mu \nu}=\left ( \begin{array}{lll} 0& \xi_{1}& \xi_{2} \cr
 \xi_{1}& m^2 &0 \cr
 \xi_{2}& 0 & n^2 \cr
  \end{array} \right ),~~~g^{\mu \nu}= {1 \over \Delta^2}\,\left ( \begin{array}{lll}
   -m^2\,n^2& \xi_{1}\,n^2& \xi_{2}\, m^2 \cr
 \xi_{1}\,n^2& \xi_{2}^2 &-\xi_{1}\,\xi_{2} \cr
 \xi_{2}\,m^2& -\xi_{1}\,\xi_{2} & \xi_{1}^2 \cr
  \end{array} \right )
\end{equation}
The equation (\ref{kil2}) reduces to

\begin{equation}\label{kil10}
\xi_{1,z}-\xi_{2,r}=2w \Delta
\end{equation}
The function $\rho$ can be calculated in terms of the metric
functions $\xi_{1}, \xi_{2}, m$ and $n$.  By using a
transformation $r=f(R,Z)$ and $z=g(R,Z)$ where $f$ and $g$ are any
differentiable functions, without loosing any generality, we can
let one of the functions $\xi_{1}$ or $\xi_{2}$ zero . Here in
this work we will take $\xi_{1}=0$ and let $\xi_{2}=q$. With these
simplifications we obtain $\Delta=q\,m,$, $q_{,r}=-2w\,q\, m$, and

\begin{equation}
ds^2=2\,q\,dz\,dt+m^2\, dr^2+n^2 dz^2.
\end{equation}
where $w$ is a nonzero constant. With such a choice we have
\begin{equation}
\rho={1 \over m^3\,q^3}\, [2w q m^2\,n\,
n_{,r}+nqm_{,r}n_{,r}-m^2qm_{,zz}+m^2m_{,z}q_{,z}-mnqn_{,rr}-mq(n_{,r})^2]
\end{equation}

\vspace{0.3cm} \noindent {\bf Remark 3}. If $w=0$, the Killing
vector becomes hypersurface orthogonal and $q$ becomes a constant.
Then
\begin{equation}
\rho={1 \over m^3\,q^2}\, [nqm_{,r}n_{,r}-m^2
m_{,zz}+m^2m_{,z}q_{,z}-mn n_{,rr}-m (n_{,r})^2].
\end{equation}

 \vspace{0.3cm}

\noindent {\bf (5.B)}. Letting $q=e^{y}$ and $n=m$ we get $m=-{1
\over 2w}\, y_{,r}$, $w \ne 0$. Here $y$ is a function of $r$ and
$z$. Then

\begin{equation}
\rho={1 \over e^{2y}\, y_{,r}}\, [-(y_{,zz}+y_{,rr})+{1 \over 2}\,
((y_{,r})^2+(y_{,z})^2)]_{,r}
\end{equation}
We get a  nonlinear partial differential equation for $y$  when
$\rho=\rho_{0}$ is a constant.

\begin{equation}\label{y1}
-(y_{,zz}+y_{,rr})+{1 \over 2}\, ((y_{,r})^2+(y_{,z})^2)={1 \over
2}\, \rho_{0}\, e^{2y}+\rho_{1}
\end{equation}
where $\rho_{1}$ is an integration constant. An exact solution of
the above equation is given as $y=-\ln [A+B (r-r_{0})^2]$ where
$\rho_{1}=0$ and $\rho_{0}=4AB$. After performing some scale
transformations $z \rightarrow \sqrt{A}\, z, r \rightarrow
\sqrt{A}\, r$ and letting $B=w$ and  $r_{0}=0$ the metric becomes

\begin{equation}\label{y2}
ds^2= {dt\,dz \over \rho_{0}+w \,r^2}+ {r^2 \over
[\rho_{0}+w\,r^2]^2}\,(dr^2+dz^2)
\end{equation}
This metric solves Einstein field equations (\ref{ein}) with null
fluid whose energy density $\rho=\rho_{0}$ is a constant. All
curvature invariants are constants. For instance $R=-6w^2$,
$R^{\alpha \beta}\, R_{\alpha \beta}=12 w^4$. Hence the spacetime
is not asymptotically flat. When $\rho_{0}=0$ the solution
corresponds to AdS$_{3}$.

 \vspace{0.3cm}

\noindent {\bf (5.C)}. When $w=0$, the Killing vector becomes
covariantly constant. By taking $m=n$ the function $\rho$ becomes

\begin{equation}
\rho=-{1 \over mq}\,(m_{zz}+m_{rr})
\end{equation}
and the metric takes the form
\begin{equation}
ds^2=2\,q\,dz\,dt+m^2\, (dr^2+dz^2).
\end{equation}
where $q$ is a constant. Einstein tensor becomes
\begin{equation}
G_{\mu \nu}=\rho \xi_{\mu}\, \xi_{\nu}
\end{equation}
All scalars constructed from the ricci tensor vanish. We will not
consider this case in the sequel.

\section{Topologically Massive  Gravity Theory}

Topologically Massive Gravity (TMG) equations found by Deser,
Jackiw and Templeton (DJT) \cite{djt}. Recently \cite{car},
\cite{song} this theory was extended to the case  with a
cosmological constant. They are given as follows.

\begin{equation}
G^{\mu}_{\nu}+{1 \over \mu}\,C^{\mu}\,_{\nu}=\lambda\,
\delta^{\mu}_{\nu}. \label{djt}
\end{equation}
Here $G_{\mu \nu}$ and $R_{\mu \nu}$ are the Einstein and Ricci
tensors respectively and $C^{\mu}_{\nu}$ is the Cotton tensor
which is given by
\begin{equation}
C^{\mu}\,_{\nu}=\eta^{\mu \beta \alpha }\,\, (R_{\nu \beta}-{1
\over 4}\,R\, g_{\nu \beta})_{;\alpha}.
\end{equation}
The constants $\mu$ and $\lambda$ are respectively the DJT
parameter and the cosmological constant. Solutions of this theory
were studied by several authors \cite{gur4}-\cite{gur5}.

\vspace{0.3cm}

\noindent {\bf 1}.\, When the spacetime admits a non-null Killing
vector field $u^{\mu}$, assuming that the ricci scalar $R$ is a
constant (or the Gaussian curvature,  $R=2K+2w^2$, of the two
spaces orthogonal to the Killing directions is a constant) and
using the Eq. (\ref{eqn27}) we get

\begin{equation}
C_{\mu \nu}=-w\,(3 w^2+{R \over 2})[g_{\mu \nu}-{3 \over \alpha}\,
u_{\mu}\, u_{\nu}].
\end{equation}
Using the TMG field equations we obtain

\begin{equation}\label{con}
\mu=3w,~~ \lambda=-{R \over 6}.
\end{equation}
which is valid for both spacelike and timelike cases. The metrics
(\ref{god1}) with equation (\ref{kil00}) and (\ref{god2}) with
equation (\ref{kil01}) are exact solutions of TMG with parameters
satisfying the conditions (\ref{con}).

\vspace{0.3cm}

\noindent
 {\bf 2}.\,If the spacetime admits a null Killing vector the
Einstein tensor takes the form (\ref{ein}). Using the Killing
equation (\ref{kil2}) and the Einstein tensor (\ref{ein}) we get

\begin{equation}
C^{\mu}\,_{\nu}=(w \rho-\sigma)\, \xi_{\mu}\, \xi_{\nu},~~
\sigma={\xi_{1}\, \rho_{,z}-\xi_{2}\, \rho_{,r} \over \Delta}
\end{equation}
which leads to the following equations

\begin{eqnarray}
&\lambda=w^2, \label{djt1}\\
 &(\mu+w) \rho=\sigma={\xi_{1}\, \rho_{,z}-\xi_{2}\, \rho_{,r}
\over \Delta}. \label{djt2}
\end{eqnarray}
In addition to the Killing equation (\ref{kil10}) these equations
constitute the field equations to be solved for the DJT Theory.
With the simplification done in {\bf 5.A} of the last section we
get

\begin{equation}
\rho=\rho_{0}\, q^{\mu+w \over 2w},~~m=-{q_{,r} \over 2wq}
\end{equation}
where $\rho_{0}$ is an arbitrary constant. As far as the solutions
are concerned we have the following three classes:

\vspace{0.3cm}

\noindent {\bf a)}.
 We obtain a simple solution of (\ref{djt2}) when $\mu=-w$ which leads to $\rho=\rho_{0}$ constant.
 Then using the simplifications done in {\bf 5.B} of last section ($\xi_{1}=0$ and $\xi_{2}=q=e^{y}$, $n=m=-{1 \over
2w}\, y_{,r}$)  we get all metric functions  related to the
function $y$ which satisfies the differential equation (\ref{y1}).
An exact solution and metric of the spacetime is given in
(\ref{y2}) of {\bf 5.B} part of the last section.

\vspace{0.3cm}

\noindent {\bf b)}. A more general solution is obtained when $\mu
+3w=0$ where the function $y$ satisfies the equation

\begin{equation}\label{y2b}
-(y_{,zz}+y_{,rr})+{1 \over 2}\, ((y_{,r})^2+(y_{,z})^2)=
\rho_{0}\,y+\rho_{2}
\end{equation}
where $\rho_{2}$ is an integration constant. A solution of this
equation is $y={\rho_{0} \over 2}\, r^2, ~\rho_{2}=-\rho_{0}$.
Then the metric becomes

\begin{equation}\label{yb2}
ds^2= e^{{\rho_{0} \over 2}\,r^2}\,dt\,dz + {\rho_{0}^2\,r^2 \over
4\, w^2}\,[dr^2 +dz^2]
\end{equation}
Here $\rho_{0} \ne 0$.

\vspace{0.3cm}

\noindent {\bf c)}. The case when $\mu+5w \ne 0$. Function $y$
satisfies the equation

\begin{equation}\label{y3}
-(y_{,zz}+y_{,rr})+{1 \over 2}\, ((y_{,r})^2+(y_{,z})^2)=
{\rho_{0} \over \epsilon}\, e^{\epsilon\, y}+\rho_{3}
\end{equation}
where $\rho_{3}$ is an integration constant and $\epsilon={\mu+5w
\over 2w}$. The circularly symmetric metric has the form

\begin{equation}
ds^2=e^y\, dt\,dz+{1 \over 4\,w^2}\, e^{\epsilon y}\, [-{2
\rho_{0} \over \epsilon \, (\epsilon-1)}+2\rho_{3}\,
e^{-\epsilon\, y}+\rho_{4}\, e^{(1-\epsilon)\,y}]\,(dr^2+dz^2)
\end{equation}
where the function $y=y(r)$ satisfies the equation

\begin{equation}\label{y4}
y_{r}=\pm\, e^{{\epsilon \over 2}\,y}\, \sqrt{-{2 \rho_{0} \over
\epsilon \, (\epsilon-1)}+2\rho_{3}\, e^{-\epsilon\, y}+\rho_{4}\,
e^{(1-\epsilon)\,y}}
\end{equation}
Here $\rho_{4}$ is also an integration constant. All other cases
will be considered later.

\vspace{0.3cm}

\noindent {\bf Remark 4}. In the case of non-null  Killing vector
field when $w=0$ Killing vector becomes hypersurface orthogonal
and the Cotton tensor vanishes. TMG field equations with
hypersurface orthogonal Killing vectors were previously
considered. With a hypersurface orthogonal Killing vector Aliev
and Nutku showed that TMG allow only flat solution, \cite{an}.
Recently Deser and Franklin in \cite{des5} considered circularly
symmetric spacetime in TMG. They showed that such symmetry
(existence of the hypersurface orthogonal  spacelike Killing
vector ${\partial \over \partial \theta}$) breaks TMG with
cosmological constant down to vanishing of Cotton and Einstein
sectors separately.

\section{A New Massive Gravity in Three Dimensions}

Recently  a new, parity-preserving theory is introduced by
Bergshoeff-Hohm-Townsend (BHT) \cite{bht} in three dimensions
which is equivalent to Pauli-Fierz massive field theory at the
linearized level. Originally this theory, known as NMG does not
contain a cosmological term in the action. Adding a cosmological
constant $\lambda$ the new massive gravity field equations (CNMG)
are given as

\begin{equation}
2m_{0}^2\, G_{\mu \nu}+K_{\mu \nu}+\lambda\, g_{\mu \nu}=0
\end{equation}
where
\begin{eqnarray}
K_{\mu \nu}&=&2 \nabla^2\, R_{\mu \nu}-{1 \over 2} (R_{;\mu
\nu}+g_{\mu \nu}\, \nabla^2 \, R)-8 R^{\rho}\,_{\mu}\, R_{\nu
\rho} \nonumber \\
&&+{9 \over 2} R \, R_{\mu \nu}  +[3 R^{\alpha \beta}\, R_{\alpha
\beta}-{13 \over 8} R^2]\, g_{\mu \nu}
\end{eqnarray}
where $m_{0}$ is relative mass parameter and $\nabla^2$ is the
Laplace-Beltrami operator. Here we used the mass parameter as
$m_{0}$ not to confuse with the metric function $m$. Solutions of
these equations have been recently studied in \cite{clem4},
\cite{bht1}

\vspace{0.3cm}

\noindent 1. When the spacetime admits a non-null Killing vector
field, assuming that the ricci scalar $R$ is a constant and using
the equation (\ref{eqn27}) we get $R=2K+2w^2$ and

\begin{eqnarray}
\nabla^2 R_{\mu \nu}&=&2 w^2\,(3 w^2+{R \over 2})[g_{\mu \nu}-{3
\over \alpha}\, u_{\mu}\, u_{\nu}], \\
R^{\mu \alpha}\, R_{\nu \alpha}&=&{1 \over \alpha}\, (3w^2+{R
\over 2})(w^2-{R \over 2})\,u^{\mu}\, u_{\nu}+(w^2+{R \over
2})^2\,
\delta^{\mu}\,_{\nu}, \\
 K_{\mu \nu} &=&-{1 \over \alpha}\,(3w^2+{R \over 2})\,(20w^2+{R
 \over 2})\,u_{\mu}\, u_{\nu}+(22w^4+{9 \over 2}\,w^2\,R \nonumber\\
 &&+{R^2 \over 8})\,g_{\mu \nu}.
\end{eqnarray}
Then using the field equations of NMG gravity we obtain ($3w^2+{R
\over 2} \ne 0$)

\begin{equation}\label{con1}
m_{0}^2=-10w^2-{R \over 4},~~\lambda=-2w^4-{R^2 \over
8}-4\,w^2\,R.
\end{equation}

\vspace{0.3cm}
 \noindent The metrics (\ref{god1}) with equation
(\ref{kil00}) and (\ref{god2}) with equation (\ref{kil01}) are
exact solutions of NMG with parameters satisfying the conditions
(\ref{con1}). As an example the metric found in \cite{gur4} as an
exact solution of TMG is also an exact solution of the NMG with
parameters satisfying the conditions (\ref{con1}).\\

If $3w^2+{R \over 2} =0$ then we obtain
\begin{equation}
\lambda+2 m_{0}^2 w^2+{1 \over 2}w^4=0
\end{equation}

\vspace{0.3cm}

\noindent
 2. When the spacetime admits a null Killing vector field, using Eq.(\ref{kil2}) and the
Einstein tensor (\ref{ein}) we obtain $R=-6w^2$ and

\begin{eqnarray}
\nabla^2\, R_{\mu \nu}&=&(\nabla^2\, \rho+4w\sigma+6w^2 \rho)\,
\xi_{\mu}\, \xi_{\nu}, \\
 R^{\rho}\,_{\mu}\, R_{\rho \nu}&=&-4w^2 \rho\, \xi_{\mu}\,
\xi_{\nu}+4w^4\, g_{\mu \nu},\\
K_{\mu \nu}&=&[2\, \nabla^2\, \rho+8w\sigma+17w^2
\rho]\,\xi_{\mu}\, \xi_{\nu}-{1 \over 2}\,w^4 \,g_{\mu \nu}.
\end{eqnarray}
These equations lead to the following results

\begin{eqnarray}
& 2\, \nabla^2 \, \rho+(2\, m_{0}^2+17 w^2 )\, \rho+8 w \sigma=0, \label{bht1}\\
 &4 m_{0}^2 w^2-w^4+2 \lambda=0. \label{bht2}
\end{eqnarray}
The full massive gravity field  equations reduce to (\ref{bht1}),
(\ref{bht2}) and the equation (\ref{kil2}). A circularly symmetric
solution  of this equation is given by $\rho=\rho_{0}\, e^{k\, y}$
where $\rho_{0}$ is a constant and $k$ satisfies the quadratic
equation $8k^2+24k+35/2-{\lambda \over w^4}=0$. The circularly
symmetric metric becomes

\begin{equation}\label{yy1}
ds^2=e^{2\,y}\, dt\,dz+{1 \over 4\,w^2}\,[{2\rho_{0} \over k-1}\,
e^{ky}+\rho_{1} e^y]\,(dr^2+dz^2)
\end{equation}
Here $y$ satisfies a similar equation like in (\ref{y4})

\begin{equation}\label{y5}
y_{,r}=\pm \, \sqrt{{2 \rho_{0} \over k-1}\, e^{ky}+\rho_{1}\,
e^{y}}
\end{equation}
where $\rho_{1}$ is an integration constant. For each value of $k$
($k_{1}$ and $k_{2}$) we have two different metrics.

BHT introduced also a more general model \cite{bht} which also
includes the topologically massive gravity as a special case.

\begin{equation}
\lambda\, m_{0}^2 g_{\mu \nu}+\alpha \, G_{\mu \nu}+ {1 \over
\mu}\, C_{\mu \nu}+{ \beta \over 2 m_{0}^2}\, K_{\mu \nu}=0,
\end{equation}
where $\lambda$, $\alpha$ and $\beta$ are dimensionless
parameters. For $m_{0} \rightarrow \infty$ and for fixed $\mu$
generalized BHT equations reduce to the DJT equations (\ref{djt}).
The solutions of these equations are given as follows:

\vspace{0.3cm}
 \noindent
 {\bf 1.}\, Non-null case:($3w^2+{R
\over 2} \ne 0$)\\
\begin{eqnarray}
-\alpha+{3w \over \mu}-{\beta \over 2m_{0}^2}\,(20 w^2+R/2)=0, \\
\lambda m_{0}^2+\alpha w^2-{w \over \mu}(3w^2+{R \over 2})+{\beta
\over 2m_{0}^2}(22w^4+{9 \over 2} w^2 R+{R^2 \over 8})=0
\end{eqnarray}
If $3w^2+{R \over 2} =0$ then
\begin{equation}
\lambda m_{0}^2+\alpha w^2+{\beta \over 4m_{0}^2}w^4=0
\end{equation}

\vspace{0.3cm}
 \noindent
 {\bf 2.}\, Null case:

\begin{eqnarray}
& {\beta \over m_{0}^2}\, \nabla^2 \rho+(\alpha+{w \over \mu}+{17
w^2 \beta \over 2m_{0}2})\, \rho+(-{1 \over \mu}+{4 \beta
w \over m_{0}^2})\sigma=0,\\
& \lambda\, m_{0}^2+\alpha w^2-{\beta \over 4m_{0}^2}\,w^4=0
\end{eqnarray}
In a similar fashion letting $\rho=\rho_{0}\, e^{k\, y}$ we get a
quadratic equation for $k$

\begin{equation}\label{y6}
4\,w^2\,{\beta \over m_{0}^2}\,k(k+1)+ 2(-{1 \over \mu}+{4 \beta w
\over m_{0}^2})\,w\,k+(\alpha+{w \over \mu}+{17 w^2 \beta \over
2m_{0}2})=0
\end{equation}
In order that the roots to exists the following condition should
be satisfied

\begin{equation}\label{yy2}
(-{1 \over \mu}+{4 \beta w \over m_{0}^2}+2\,{\beta^2 \over
m_{0}^2})^2-4\,{\beta \over m_{0}^2}\,(\alpha+{w \over \mu}+{17
w^2 \beta \over 2m_{0}2})\,w^2 \ge 0
\end{equation}
The metric function $y$ satisfies exactly the same equation
(\ref{y5}) but $k$ solves the quadratic equation (\ref{y6}) in
this case. The circularly symmetric metric is of the form given
(\ref{yy1}) where $y(r)$ satisfies (\ref{y5}).

 Both for null and non-null cases when the scalars of the Killing
vector fields are constants the we have a more general result. The
following conjecture implies that the corresponding metrics solve
all higher derivative gravitational field equations in three
dimensions.

\vspace{0.3cm} \noindent {\bf Conjecture}. {\it Let a three
dimensional spacetime admit a Killing vector field $u_{\mu}$
(non-null or null) with constant scalars. Let the Gaussian
curvature $K$ of the two dimensional spaces be constant for the
case of non-null vector fields where corresponding Einstein
tensors are respectively given in (\ref{eqn270}) and (\ref{ein}).
Then any symmetric second rank covariant tensor constructed from
the ricci tensor by covariant differentiation and by contraction
is the linear sum of $u_{\mu}\, u_{\nu}$ and the metric tensor
$g_{\mu \nu}$}

\vspace{0.3cm}

\section{Conclusion}
In this work we first studied the Killing vector fields in a three
dimensional spacetime geometry. We showed that, independent of the
type of the Killing vector fields, the ricci tensor can be
determined in terms of the Killing vector fields and their
scalars. Usually components of this tensor are calculated in terms
of  the components of metric tensor in a  given coordinate system.
In three dimensions, when the geometry admits at least one Killing
vector field we don't have to follow such a direction to determine
the components of the ricci tensor. Using this property in each
case (when the Killing vector field is timelike, spacelike or
null) we first presented solutions of Einstein field equations
with sources. Then by using special cases, when the scalars of the
Killing vector fields are constants, we gave solutions of the
field equations of the Topologically Massive Gravity and New
Massive Gravity theories                                . Some of
the solutions of the Topologically Massive Gravity field equations
we obtained in this work may already be known. Our basic purpose
in this work is to present a new method to solve higher derivative
gravity theories rather finding specific solutions. We conjecture
at the end that, for the three type of Killing vector fields with
constant scalars our method solves all higher derivative theories
in three dimensions.

 \vspace{2cm}

This work is partially supported by the Scientific and
Technological Research Council of Turkey (TUBITAK) and  Turkish
Academy of Sciences (TUBA).

\end{document}